\documentclass[a4paper,11pt]{article}
\usepackage[utf8]{inputenc}
\usepackage{graphicx}
\usepackage{subfigure}
\usepackage{amsmath}
\usepackage{cite}
\usepackage{wrapfig}

\title{Risk and Return models for Equity Markets and Implied Equity Risk Premium}
\author{Enzo Busseti}
\date{May 19, 2010}

\begin{document}

\maketitle

\begin{abstract}
Equity risk premium is a central component of every risk and return model in finance and a key input to estimate costs of equity and capital in both corporate finance and valuation. 

The article by Damodaran \cite{damodaran} examines three broad approaches for estimating the equity risk premium.
The first is survey based, it consists in asking common investors or big players like pension fund managers what they require as a premium to invest in equity. 
The second is to look at the premia earned historically by investing in stocks, as opposed to riskfree investments. 
The third method tries to extrapolate a market-consensus on equity risk premium (Implied Equity Risk Premium) by analysing equity prices on the market today. 

After having introduced some basic concepts and models, I'll briefly explain the pluses and minuses of the first two methods, and analyse more deeply the third. In the end I'll show the results of my estimation of ERP on real data, using variants of the Implied ERP (third)  method. 
\end{abstract}

\section{Introduction}
\subsection{Equity Risk Premium}
Risk aversion is a central notion in modern finance. Put in simple terms, investors require higher returns for risky investments like stocks than safe ones like Treasuries (Federal Reserve bills and bonds). The difference is called ``risk premium'': it is the excess return expected from the risky investment above a risk-free return. In the case of stocks in general this difference is called the Equity Risk Premium. It refers to the perfomance of the overall market (in practice a  broad market index like the S\&P 500).
There is vigorous debate among experts about the method employed to calculate the equity premium and, of course, the resulting answer. 

A good estimate of equity risk premium is a key input in both corporate finance evaluation and asset management. A firm invests in new assets and capacity only if its managers think they can generate higher return than the cost of the required capital. If equity risk premium increases, the cost of raising capital increases, and thus we expect less investment in the overall economy.

In asset management, equity risk premium is implicitly used when estimating the expected return of the investments in stocks. This determines, for example, the amount of money invested in the equity market by governments and corporations to meet future pension fund or healthcare obligations.

In the following, we'll keep it simple and sidestep a few technical issues. We'll work with expected returns that are long-term, and pre-tax. 
By long-term, we mean something like 10 years, as short horizons raise questions of market timing. (That is, it is understood that markets will be over or under-valued in the short run.) 
Moreover, although individual investors should care about after-tax returns, it is convenient to refer to pre-tax returns as do virtually all academic studies. Transaction costs and all other ``market frictions'' are also neglected.

\subsection{Portfolio Theory}

In designing a portfolio, investors seek to maximize the expected return from their investment, given some level of risk they are willing to accept. 
To a first approximation, an asset's return is modelled as a normally distributed random variable, and risk defined as its standard deviation. 

A portfolio is a weighted combination of assets, so that its return is the weighted combination of the returns of these  assets. By combining different assets whose returns are not correlated, investors can reduce the total variance of the portfolio. 

Theoretically, if one could find sufficient securities with uncorrelated returns, he could reduce portfolio risk at any level he wants (by the Central Limit Theorem). Unfortunately, this situation is not typical in real financial markets, where returns are positively correlated to a considerable degree because they tend to respond to the same set of influences. We can proceed further, and roughly simplify that set of influences to a single factor, to which stock's return are correlated: the market return (as measured by a market index like S\&P 500). 

Risk for individual stock returns is then split in two components: systematic and unsystematic risk. The first is the risk involving the whole market, and cannot be reduced through diversification. Business cycles, changes in interest rate and wars all cause undiversifiable risks. 
Unsystematic risk is specific to a single stock or group of stocks. It is not correlated with general market movements: it can be for example a sudden strike by the employees of a company or some technological innovation that moves value from one company (whose business becomes outdated) to another.

To quantify the systematic risk of a stock we divide its return $R$ in two parts: one perfectly correlated with and proportional to market return $R_m$, and a second independent from the market (for simplicity a normally distributed variable $\epsilon$). Thus  we have:
\begin{equation}
R = \beta R_m + \epsilon.
\label{return}
\end{equation}
 The proportionality factor $\beta$ is a market sensitivity index, indicating how sensitive the security is to changes in the market level. It is usually estimated by linear regression on historical data (security returns versus market returns). 
 All $\epsilon$ from different stocks are uncorrelated (independent) by the above assumption, the only source of correlation among stock returns is the general market movement $R_m$. In addition, the $\epsilon$ mean sould be $0$, as it represents random umpredictable events. 

Using this definition of security return (``market model'') the systematic risk is $\beta$ times the standard deviation of the market return $\sigma_m$, and the unsystematic risk is simply the std. dev. of the residual return factor $\sigma_{\epsilon}$. 

We can compute a portfolio's $\beta_p$ factor as an average of the components' $\beta$ weighted by the proportion of each security. The portfolio's systematic risk is in turn its $\beta_p$ factor times $\sigma_m$. 
Given the assumptions we made on $\epsilon$, the Central Limit Theorem assures that the portfolio's unsystematic risk tends  instead to $0$ as the number of securities held in the portfolio grows.

\subsection{Capital Asset Pricing Model}

In the above section two measures of risk for a security have been developed: the total risk (standard deviation of return) and the relative index of undiversifiable risk $\beta$. In the capital asset pricing model (CAPM) only $\beta$ is used to quantify risk, neglecting the portion of risk that is diversifiable. Securities with higher systematic risk should have higher expected returns. 

The basic postulate underlying CAPM theory is that assets with the same systematic risk should have the same expected rate of return: that is, the prices of assets in the capital markets should adjust until equivalent risk assets have identical expected returns. 

Consider for example an investor who holds a risky portfolio with $\beta = 1$ (same risk as the market portfolio): he should expect the same return as that of the market portfolio. Another investor holding a riskless portfolio ($\beta = 0$) should instead expect the same rate of return of riskless assets, such as Federal Reserve Treasury Bills. 

Now, consider the case of an investor who holds a mixture of these two portfolios. The fraction of money allocated in the ``market'' portfolio equals the $\beta_p$ of the whole portfolio. We can also allow the investor to borrow at the risk-free rate and invest the proceeds in the risky portfolio, so that the portofolio $\beta$ becomes larger than $1$.

The expected return of the composite portfolio $R_p$ is a weighted average of the expected returns on the two portfolios, one with risk-free return $R_f$ and the other with market return $R_m$:
$$E(R_p) = (1 - \beta_p )\   R_f + \beta_p \  E(R_m)$$
or
\begin{equation}
E(R_p) =  R_f + \beta_p\  [ E(R_m) - R_f]
\label{CAPM}
\end{equation}
We can rewrite this equation in term of risk premia, obtained subtracting the risk-free rate from the rate of return. The portfolio risk premium  and market risk premium (ERP) are given by:
$$r_p = E(R_p) - R_f$$
$$r_m = E(R_m) - R_f$$
Substituting into equation (\ref{CAPM}):
\begin{equation}
r_p = \beta_p r_m.
\label{CAPMrp}
\end{equation}
In this form, CAPM states that the risk premium for a portfolio should be equal to the \emph{quantity of risk} measured by its $\beta_p$, times the \emph{market price for risk}, or Equity Risk Premium $r_p$. 

\subsubsection{Beyond CAPM}
As written above, one basic assumption of CAPM theory is that the only agent correlating different stock's returns is the overall market movement. In this simplified picture a stock's performance is linked to the surrounding economy only through its $\beta$ factor, while it's intuitively clear that many other agents play important roles in determining a stock's return. 
Various flavours of multifactor CAPM have been developed, basically extending equation (\ref{return}) to relate a stock's return to many external factors:
\begin{equation}
R =  \beta_{1}F_1 + \beta_{2}F_2 + \cdots + \beta_{n}F_n + \epsilon.
\end{equation}
The $F_i$ can be chosen among macroeconomic indexes like unemployment, inflation (or better surprises in inflation), interest rates, consumer confidence, etc. In practice every analyst chooses his own set of $F_i $ factors, then a linear regression is used to estimate the $\beta_i$. The next move is to assign a risk premium $r_i$ to each factor $F_i$, and so we can write an equation analogous to (\ref{CAPMrp}) for the stock's risk premium $r_s$:
\begin{equation}
r_s =\sum_{i =1}^n \beta_i r_i.
\end{equation}

It should be emphasized that while these models work well when dealing with a single stock perfomance, the simple CAPM gives good estimate for a broad market portfolio. The equity risk premium that appears in (\ref{CAPMrp}) is a market-wide number, affecting expected returns on all risky investments (the \emph{market price for risk}). Using a larger equity risk premium will increase the expected returns for all risky investments, and by extension, reduce their price. 

\subsection{Equity Risk Premium Determinants}
Economic risk, or  the health and predictability of the overall economy, is of great importance. Put in simpler terms, the equity risk premium should be lower in an economy with predictable inflation, interest rates and economic growth than in one where these variables are volatile.

The risk aversion of investors in the markets is a critical factor in ERP evaluation. Higher risk aversion means higher risk premia, and lower equity prices, on the contrary as risk aversion declines, risk premia will fall. 
It is the collective risk aversion of investors (across the whole market), that influences equity risk premium. Different groups may have very different risk aversion profiles (for example it is understood that risk aversion grows with age).

The relationship between information and equity risk premium is complex. More precise information should lead to lower risk premia, since it becomes simpler to forecast future earnings and cash flows. On the other hand, in general more informations (possibly unprecise) create more uncertainty about future earnings, since investors disagree about how best to interpret these.
In addition, information differences may be one reason why investors demand larger risk premia in some emerging markets than in others: markets vary widely in terms of transparency and information disclosure requirements.

\section{Survey Based Evaluation}
One method to evaluate equity risk premium is to ask investors what they require as excess return from investments in stock over the risk free rate. Since there are millions of investors in the stock market, one challenge is to find a representative subset. These are some of surveys used in practice.
\begin{itemize}
\item Individual investors are polled about their optimism for future stock prices, and their expected return on stocks. These polls are too much sensible to recent stock price movements, showing higher investors' confidence after periods of price growth. Psychological effects play also an important role, as it seems that survey numbers vary depending on the framing of the questions.
\item Institutional investors and financial professionals are also surveyed. Some services track financial newletters and forums to estimate an advisor sentiment index about the future direction of equities. Others ask the Chief Financial Officers (CFOs) of companies about their expected equity risk premium. These surveys provide values which make more sense than the indivduals', but still they show too big standard deviations to be taken seriously. Expected premia vary from 1.2\% at the first percentile to 12.4\% at the tenth.
\item Some surveys focus on academics, asking financial economists what they think is a good estimate of equity risk premium. The rationale here is that economists' opinion is highly influential to both students (who are the future managers) and practitioners, who read papers and books.
As with the other survey estimates, there is a wide range of opinion (too high standard deviation) and in general the values are higher than those proposed by institutional investors.
\end{itemize}
This method shows therefore serious drawbacks: results vary wildly depending on the set of investors or experts chosen, and even choosing restricted sets of people with similar market influence and knowledge does not help.

\section{Historical Premium Evaluation}
This is probably the most widely used approach to estimate equity risk premium: the returns earned on stocks over a long time period are estimated, and compared to the returns of a default-risk-free asset on the same time period. Again there are large differences between the various estimates: here are some reasons for these divergences.
\begin{itemize}
\item Different time periods used for estimation, i.e. how far back in time we should go to estimate this premium (in the U.S. market data are available from $\approx 1870$).
Using shorter and recent time periods has the advantage of providing an up-to-date estimate, as investors' and market features have changed much over the last century (even in mature markets like the U.S.). On the other hand a longer time period means more points to average on, and therefore lower noise (standard deviation) in the risk premium estimate. 
A good compromise might be to use an exponential weighted average that gives more weight to recent years, and progressively less to the points further back in time.

\item There are many possible choices of risk-free rates. We can choose as risk-free rate returns earned by every type of treasury security: short-term (treasury bills in the U.S.) or long term (treasury bonds). The logic choice here are the treasuries whose time to maturity is most similar to the time horizon on which we are interested. For a medium-long term investment, if we want to avoid problems of market timing, a ten year treasury bond is a good choice. 

\item There are many possible way to measure equity returns, too. A natural choice is a popular market index (like Dow Jones), since its historical values are more widely accessible and reliable than individual stocks'. It should be noted however, that if we are interested in the perfomance of the whole stock market we should measure returns on the broadest possible portfolio (Dow Jones consists of only the 30 biggest companies) so other indices with shorter history, like S\&P 500, may be better. A common error that can be caused by the choice of a narrow index is the so called ``survivor bias'', because companies which went bankrupt are singled out from the index calculation: the returns may be biased upwards. 

\item There is some debate about whether it's better to use nominal returns or real returns (adjusted for inflation). As long as equity risk premium is concerned however, we should not bother: inflation rate would add to both risk-free and equity returns, so after taking the difference it vanishes. 

\item When computing the average risk premium on a given period, for example over 10 years starting from a series of returns for each quarter, we can opt for a geometric or an arithmetic average. This is an open debate, a study by Indro and Lee \cite{IndroLee} in 1997 concludes that both approaches are biased (the arithmetic upwards, and the geometric downwards).
A well known inequality states that $M_{arithmetic} \geq M_{geometric}$, and the two are equal if and only if all returns averaged are equal. Being the returns very volatile, the difference between the results of the two averaging approaches can be substantial.
Again, many practitionists use a compromise of the two, a weighted average of arithmetic and geometric mean (see \cite{Blume}).

\end{itemize}

In Table \ref{historicalERP}, taken from \cite{damodaran}, we can see how important are the differences due to these choices. The author used raw returns data for stock, 6-month treasury bill and ten-year treasury bond, from 1928 to 2008.

\begin{table}[htbp]
\begin{center}
\begin{tabular}{|c||c|c|c|c|}
\hline
 &   \multicolumn{2}{|c|}{   ERP: Stocks minus T.Bills} & \multicolumn{2}{|c|}{ ERP: Stocks minus T.Bonds}\\
\hline
         & Arithmetic  &    Geometric & Arithmetic  &     Geometric\\
\hline
1928-2008  &  7.30\%&          5.65\%&      5.32\%&             3.88\%\\
\hline
1967-2008  &  5.14\%&         3.33\% &     3.77\%&            2.29\%\\
\hline
1997-2008  & -2.52\%&       -6.26\%  &   -4.52\% &           -7.95\% \\
\hline
\end{tabular}
\caption{Overview of historical Equity Risk Premium (ERP) calculation results with different estimation periods, riskfree rates and averaging approach. From the article \cite{damodaran}.}
\label{historicalERP}
\end{center}
\end{table}

\section{Implied Equity Risk Premium}
Another method widely used in estimates consists in extrapolating required rates of return from the market prices of equities. Then, subtracting today's risk-free rates, we have an up-to-date estimate of equity risk premium.

The idea is based on the basic concept that as investors price an assets, they are implicitly telling what they require as an expected return on it. Price is defined as the present value of all asset's expected future cash flows (\emph{Discounted Cash Flow Model}):

\begin{equation}
 P = \sum_{i = 0}^n \frac{CF_{i}}{(r) ^ i}
\label{prezzo}
\end{equation}

where $P$ is the price of the security now, $CF_{i}$ is the expected cash flow at the end of period $i$, and $r$ the required yield on each period.

For an investment in equity the expected cash flows are dividends (one for each period) and the money received at the end when selling back.

\subsection{Dividend Based Approach (Gordon Model)}
This model determines the fair value of a stock as the present value of the series of its dividend payments, assuming that dividends will grow at a constant rate.
In practice, the fair price $P$ of a stock is a function of the dividend payed out today $D$, its expected growth rate $g$ and the required rate of return $k$:
\begin{equation}
P= \sum_{t=1}^{\infty}  D\times\frac{(1+g)^t}{(1+k)^t} = \frac{D_1}{k-g}
\label{gordon}
\end{equation}
where $D_1 = D (1 + g)$ is the dividend next period. Solving for $k$ we obtain:
$$k = \frac{D_1}{P} + g$$
so in this model the required rate of return on a stock is simply its dividend yield plus the expected growth rate of its dividend payouts.

This model assumes that the firm in evaluation will sustain constant growth forever. This is not an acceptable simplification in many practical cases: for example we should allow earnings to grow at extraordinary rates for the short term, and then settle at the risk-free rate (as extrapolated from the term structure of Fed. bonds interest rates). 
So we would split the sum in Eq. (\ref{gordon}) in different parts for the short and long term dividends/growth. 

\subsection{Earnings Based Approach}
This is a slightly more refined variant of the dividend model, where we focus on earnings instead of dividends. We are addressing a flaw in the Gordon Model, which assumes that companies pay out as much as they can afford to in dividends (and this is generally not true).  We make some substitution in Eq. (\ref{gordon}): 
\begin{itemize}
\item we replace $D_1$ with the Earnings Per Share (EPS) times the payout ratio $p$ (the fraction of earnings distributed in dividends);
\item the expected growth rate of dividends $g$ is substituted with the expected rate of return $k$ times the retention ratio (fraction of earnings reinvested, $1-p$). The rationale is that the growth of earnings is due to the fraction of earnings reinvested, at a rate equal to the expected rate $k$ (of market consensus).
\end{itemize}
\begin{equation}
P = \frac{EPS \cdot p}{k - k\cdot(1-p)} = \frac{EPS}{k}
\label{EPSapproach}
\end{equation}
The inverse of the PE ratio (also referenced as the earnings yield) becomes the required return on equity.

This model too has the potential flaw of assuming constant growth of the stock. This assumption has two components: constant dividend payout ratio and constant return on investments. The former is not true in the long run, especially for younger firms who reinvest all earnings in the first ``growth'' years and then start paying dividends. The latter is also questionable, both because growth rates change with time (a correction similar to that discussed above may help), and because managers may not be able to invest all earnings at the same rate $k$.

\section{Analysis}
I have used the last method described to extrapolate an implied equity risk premium from the market value of the stocks' prices. 
My goal is to see how the formula (\ref{EPSapproach}), for the estimation of equity risk premium, behaves with daily data. 

\begin{figure}[htbp]
\begin{center}
\includegraphics[scale=0.33]{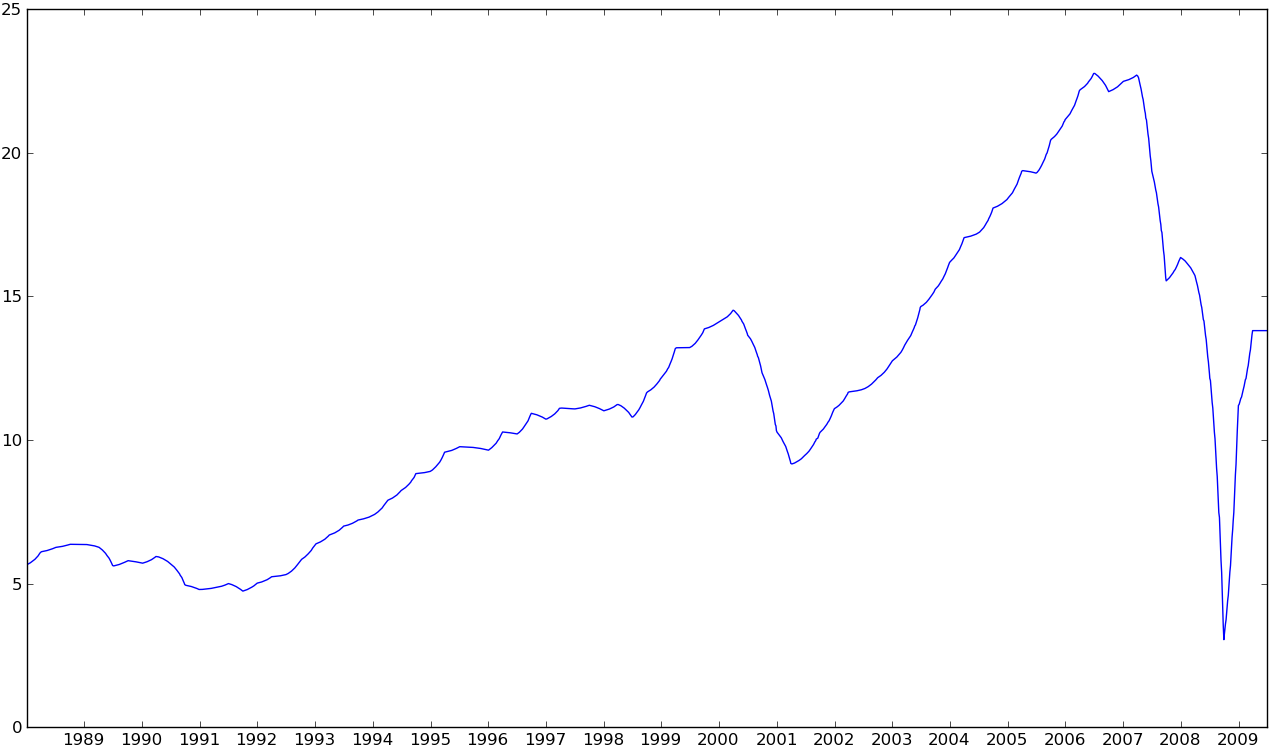}
\caption{Moving average of EPS quarterly estimation for all firms in S\&P 500 index, values are in dollars.}
\label{EPSestimation}
\end{center}
\end{figure}

I worked with the S\&P 500 index: from their site I got EPS estimation (weighted average for all firms composing the index) for each quarter, from 1988 to 2009. I smoothed the data with an exponential moving average of period 50 days (less than the sampling frequency of $\approx 100$ days). The result is shown in Figure \ref{EPSestimation}.

\begin{figure}[htbp]
\begin{center}
\includegraphics[scale=0.33]{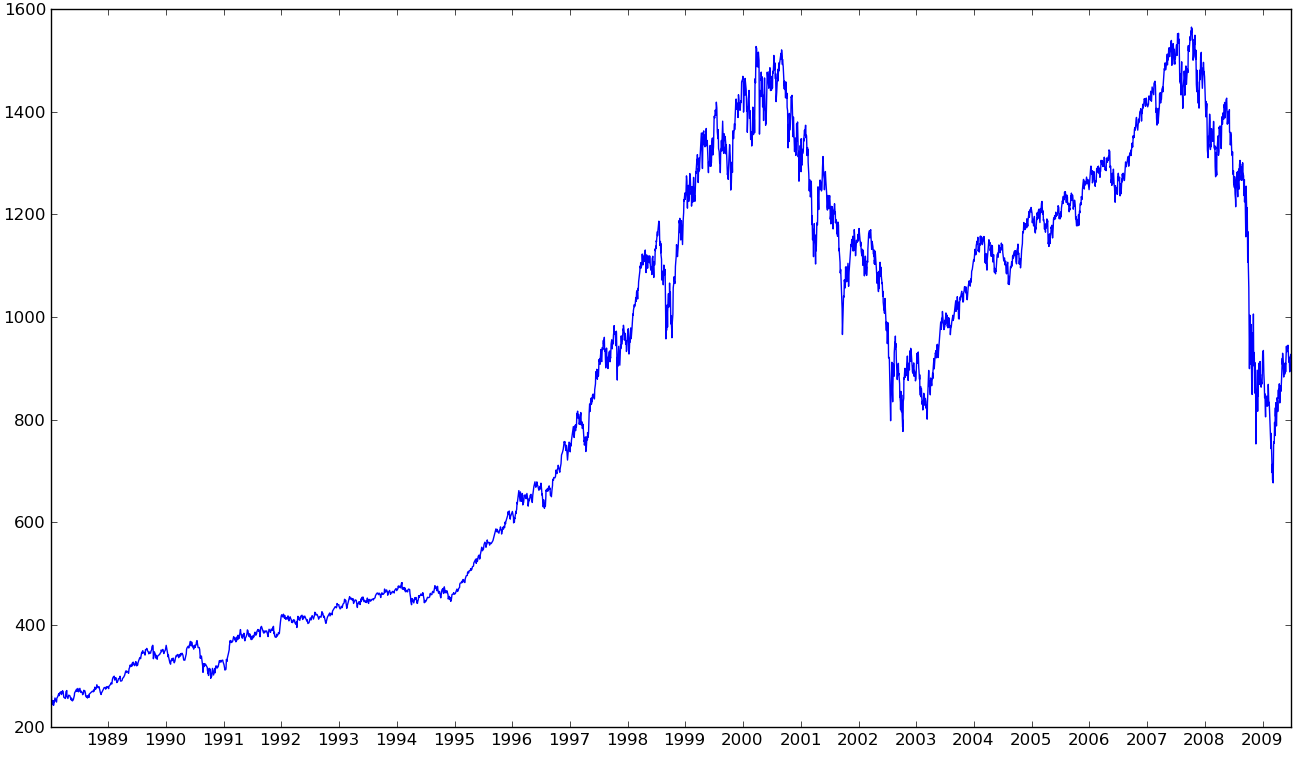}
\caption{Daily values of the S\&P 500 index, in dollars.}
\label{priceArray}
\end{center}
\end{figure}

 Then I downloaded the daily values of the index in those twenty years, reported in Figure \ref{priceArray}.

\begin{figure}[htbp]
\begin{center}
\includegraphics[scale=0.33]{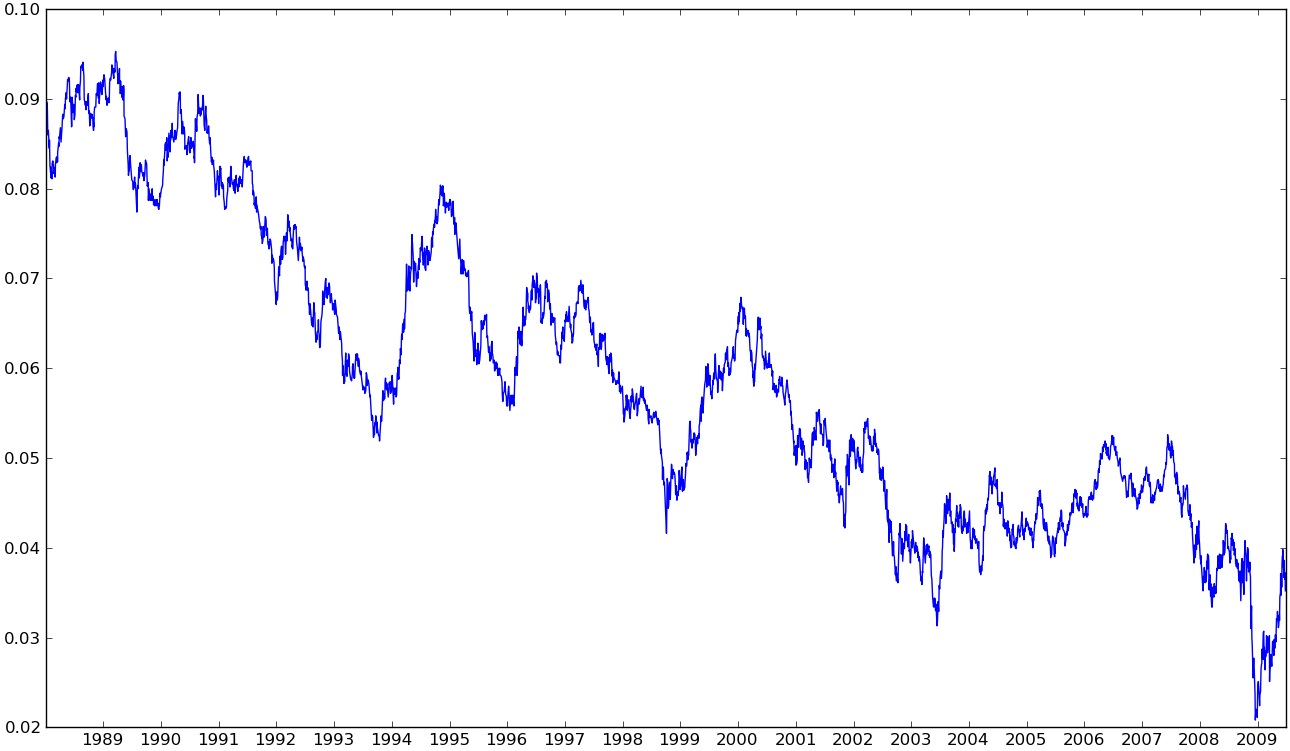}
\caption{Daily quotations of the annual yield for 10 years federal bonds.}
\label{ratesArray}
\end{center}
\end{figure}

Finally I got from the site of the Federal Reserve daily quotations of bonds rates. In Figure \ref{ratesArray} we have annual yields for 10 years bonds.

\begin{figure}[htbp]
\begin{center}
\includegraphics[scale=0.15]{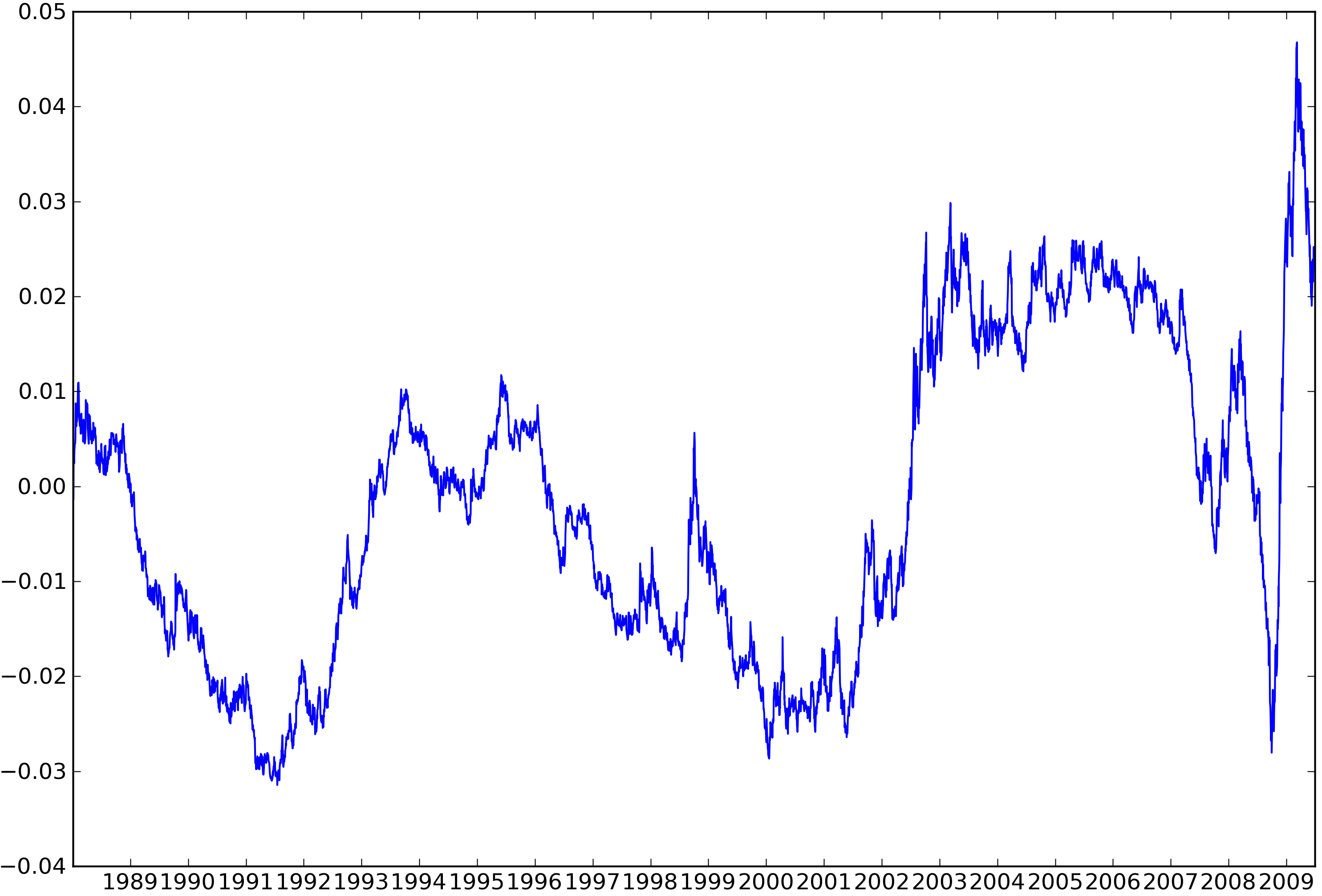}
\caption{Daily estimations of the Equity Risk Premium, obtained with formula (\ref{EPSapproach}) and the data shown in the preceding figures.}
\label{ERPestimated}
\end{center}
\end{figure}

The next step was to apply the formula (\ref{EPSapproach}) discussed above, and obtain its estimation of Equity Risk Premium. The result is shown in Figure \ref{ERPestimated}.

The first striking result we notice is that the premium estimated in this way is not always positive, there were two long periods (1989-1994 and 1996-2002) when it went below zero. We can interpret these as investors' expectations of greater perfomance than a simple growth of earnings. Then, the 2008 crisis shows a sudden decrease in risk premium, as EPS went down, and then the expected big increase when prices fell.

In general we can say that Eq. (\ref{EPSapproach}) is still too na\"if to provide reliable result over a wide range of data. There's only a period (2003-2007) in which the resulting premium has meaningful and almost constant value, otherwise it oscillates with the movement of the underlying variables, showing that there must be hidden complexity ignored by Eq. (\ref{EPSapproach}).

\end{document}